\documentclass[12pt]{article}
\usepackage{graphicx}
\font\tenrm=cmr10

\newcommand{\delR} {\delta\!R}
\newcommand{\simle} {\mathrel \rlap {\raise 0.511ex \hbox{$<$}{\lower 0.511ex
\hbox{$\sim$}}}}

\begin{document}
% new macro for bibliography
\renewenvironment{thebibliography}[1]
  { \begin{list}{\arabic{enumi}.}
    {\usecounter{enumi} \setlength{\parsep}{0pt}
     \setlength{\itemsep}{3pt} \settowidth{\labelwidth}{#1.}
     \sloppy
    }}{\end{list}}

\parindent=1.5pc
\vglue -2.0cm
\hfill
TAUP 2000-92
\vglue 2.cm
\begin{center} {INSTABILITY OF BUBBLES \\
               \vglue 3pt
               NEAR THE HADRON--QUARK-GLUON-PLASMA PHASE TRANSITION}\\
\vglue 5pt
\vglue 1.0cm
{Gideon Lana }\\
\vglue 0.2cm
{and}\\
\vglue 0.2cm
{Benjamin Svetitsky}\\
\vglue 0.5cm
\baselineskip=14pt
{\it School of Physics and Astronomy}\\
\baselineskip=14pt
{\it Raymond and Beverly Sackler Faculty of Exact Sciences}\\
\baselineskip=14pt
{\it Tel Aviv University\\69978 Tel Aviv, Israel}\\

\vglue 0.8cm
{ABSTRACT}
\end{center}
\
\footnotetext[0]{Presented at the Workshop on "QCD Vacuum Structure and its
Applications", June 1-5 1992, Paris, France.}
\vglue 0.3cm
{\rightskip=3pc
 \leftskip=3pc
 \tenrm\baselineskip=12pt
 \noindent
The small surface tension of the interface between hadronic and
quark-gluon-plasma domains, along with a negative curvature tension, implies
that the uniform plasma is unstable against spontaneous formation of hadronic
bubbles. We furthermore show that spherical bubbles are not stable against
small perturbations.}

\vglue 0.8cm
{\bf\noindent 1. Introduction}
\vglue 0.2cm

\normalsize
\baselineskip=14pt

In what follows, we shall adopt the common assumption that the
quark-gluon-plasma phase turns into the hadron phase via a first order
transition. In other words, that at some temperatures the two phases may
coexist.

First order phase transition usually proceeds by nucleation of bubbles.
Consider for example, boiling water at atmospheric pressure near the transition
temperature $T_0=37$~K. The free energy of a bubble of vapor in the water has
a free energy which depends on its radius:

$$
F(R)=\Delta P \, {4\over 3}\pi R^3 + \sigma \, {4\pi R^2}\ +
 \ldots \ . \eqno{(1)}
$$
$\Delta P$ is the difference between the pressures of the two phases (zero at
the transition temperature $T=T_0 \ $, and positive above); $\sigma$ is the
excess of free energy per unit area for a planar interface.

$F(R)$ is shown schematically in Fig.~1.
For $T<T_0$, both the volume and surface terms act to suppress bubble
formation;
any bubble formed by a fluctuation collapses to $R=0$.
For $T>T_0$, however, the volume term  $\Delta P\,4
\pi R^3 / 3 $ encourages growth of bubbles
while the surface term  $\sigma \, {4\pi R^2}$ creates a barrier.
The maximum of $F(R)$ occurs at some $R=R_c$.
A bubble formed in the superheated liquid with $R<R_c$ will shrink
away, but a bubble formed with $R>R_c$ will grow until the entire
system is vapor.

\begin{figure}[ht]
\begin{center}
\includegraphics*[width=.8\columnwidth]{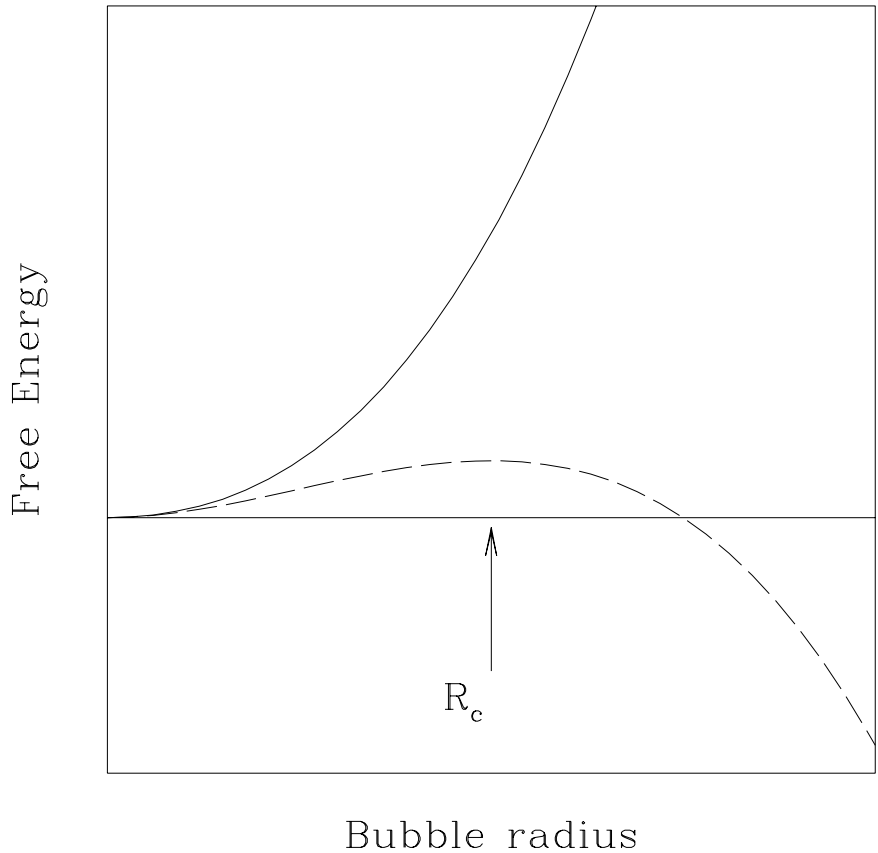}
\caption{ \baselineskip=12pt
{\tenrm Schematic representation of bubble free energy in boiling water,
retaining only volume and area terms, for $T<T_0$ (solid) and $T>T_0$
(dashed).} }
\end{center}
\end{figure}

In Eq.~(1) we have indicated that there might be more terms beyond
the area term, having in mind an expansion in powers of $1/R$.
In  condensed-matter contexts, a ``curvature'' term, $\alpha\cdot 8\pi R$, is
clearly negligible; by dimensional analysis, $\alpha/\sigma \sim \lambda$,
where
$\lambda$
is a microscopic scale, and  $(\alpha R)/(\sigma R^2) \sim
\lambda/R$,
an exceedingly small number since the typical nucleating bubble is
much larger than either the inter-molecular spacing or the
correlation length which are usually of the order of angstroms.
The picture in Fig.~1, based on volume and area terms, is then the
whole story.

Note that our picture will be substantially the same for liquid
droplets in the vapor as for vapor bubbles in the liquid, since
$\sigma$ is the same in the two cases.
This is because the surface tension may be measured in planar
interfaces, and it doesn't matter which phase is on which side.

\vglue 0.8cm
{\bf\noindent 2. The QCD Transition}
\vglue 0.2cm

In QCD (at least in the limit of massless quarks) there is only one scale.
Masses, radii, the transition temperature, the latent heat, etc.
are all on the order of 200 MeV or 1 fm.
This is then the only ``microscopic'' size, and 1 fm should also
be the typical size of a bubble, and one cannot ignore additional terms in (1).
Indeed, Monte Carlo measurements $^{2,5,7}$ and MIT bag model
calculations$^{1,3}$ found that for QCD  $\sigma/T_0^3 $ turns out to be a
small number, and one is forced to include additional terms,
$$
F(R)=\Delta P \, {4\over3}\pi R^3 +  \sigma \, {4\pi R^2}\ +
\alpha \, 8\pi R + \ldots\ .
\eqno(2)
$$

 Furthermore, it was found$^3$ that $\alpha$ is sizeable and negative.
Fig.~2 depicts $F(R)$ for a hadron bubble within the quark gluon plasma. It is
qualitatively different than that shown in Fig.~1. Indeed, the minimum in the
free energy for $T>T_0$ implies that vacuum bubbles are stable.  The minimum of
$F(R)$ is not a mere artefact of using the expansion (2): an exact bag
model calculation$^3$ exhibits a very similar behavior.

 \begin{figure}
\begin{center}
\includegraphics*[width=.8\columnwidth]{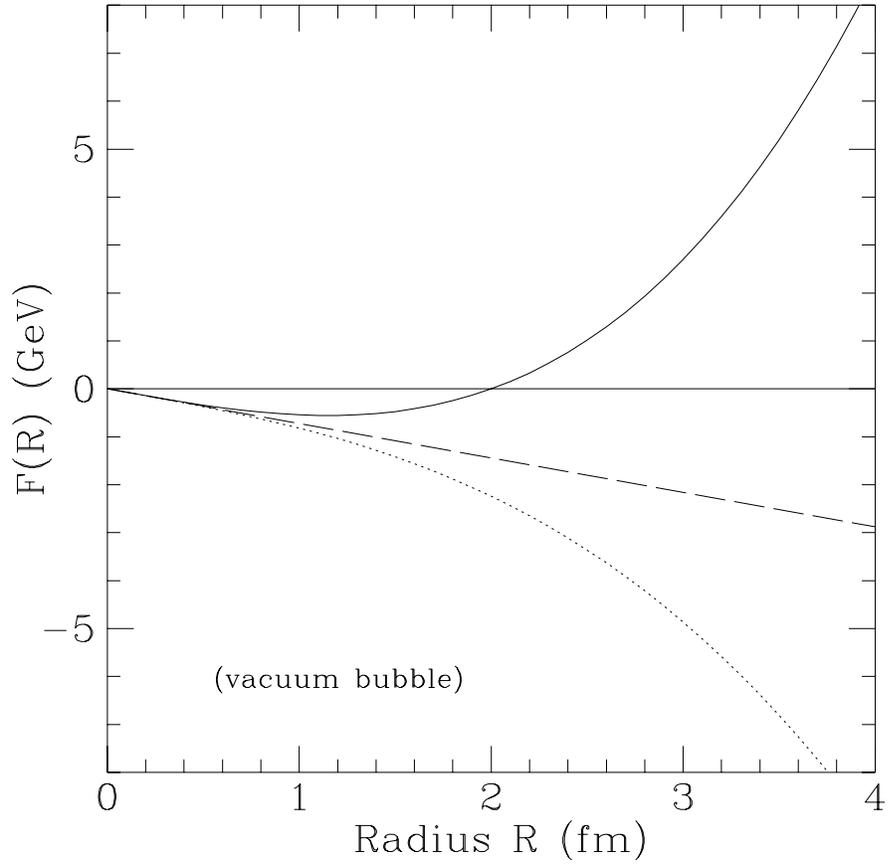}
\caption{\baselineskip=12pt
{%\tenrm
 Free energy of a hadron gas bubble in the plasma for temperatures near  the
transition temperature, fixed at $T_0=150~MeV$. $T=155~MeV$ (solid),
$T=T_0=150~MeV$ (dashed) and $T=147~MeV$ (dotted).  The figures are taken from
a bag model calculation of Ref.~3. } }
\end{center}
\end{figure}

It was further suggested$^{3,4}$ that this result might transcend the bag
model:
All that is required is a thin interface, the energy of which is due
primarily to the reaction of the fields on either side.
A lattice calculation of $\alpha$ in the pure glue theory$^5$ apparently
supports this picture.

The fact that one can lower the free energy of the plasma phase by
creating a vacuum bubble seems at first glance to imply that the
equilibrium plasma is a Swiss cheese of bubbles, the density of which
is determined by the (unknown) bubble--bubble interaction.
This would only be true, however, if (1) the bubbles do not affect
each other, and (2) each bubble is stable in itself.
A simple calculation will teach us, however, that
spherical bubbles are unstable against deformation, and that growth
of long, narrow structures is favored.
We do not address the problem of determining the actual equilibrium
configuration of the high-temperature phase, a problem which will
require much more sophisticated analysis.

In order to study the stability of the spherical bubble against small
perturbations, we expand the free energy around the equilibrium spherical
bubble, of radius $R_0$.
Minimizing (2), we find that $R_0$ satisfies
$$
4\pi\Delta P\,R_0^2 + 8\pi\sigma R_0 + 8\pi\alpha=0.
\eqno{(3)}
$$
The generalization of (2) to non-spherical shapes is$^6$
$$
F=\Delta P \,V + \sigma\int dS +\alpha\int dS\,\left({1\over
R_1}+{1\over R_2}\right) + \cdots,
\eqno{(4)}
$$
where the integrand in the third term is the local extrinsic curvature,
with $R_1$ and $R_2$ representing the principal radii of curvature.
Let the bubble surface be specified by
$$
R(\theta,\phi)=R_0 + \delR(\theta,\phi).
\eqno{(5)}
$$

The total $O(\delR)$ contribution to $F$ vanishes identically on the
equilibrium radius $R=R_0$.
The correction to the free energy is then second order and may be written as
$$
\delta F = \int \sin \theta\, d \theta \,d\phi \,\delR\left( A + B L^2 \right)
\delR.\eqno{(6)}
$$
$-L^2$ is just the angular part of the Laplacian operator, and
$$ A=\Delta P\, R_0 + \sigma,
\qquad B={\sigma\over 2}+{\alpha\over R_0} . \eqno{(7)}
$$
Recall that we presume $\alpha<0$.
When $R_0$ is determined by (3), it is easy to see that
$A$ is positive and $B$ negative.

Expanding $\delR$ in a multipole expansion,
$$
\delR(\theta,\phi)=\sum_{l,m}c_l^m Y_{lm}(\theta,\phi),\eqno{(8)}
$$
we have
$$
\delta F = \sum_{l,m}[A+Bl(l+1)]\left|c_l^m\right|^2\ .\eqno{(9)}
$$
Evidently $\delta F>0$ for a monopole perturbation, since (2) has been
minimized;
for a dipole perturbation we obtain $\delta F=0$ , consistent
with the fact that $l=1$ corresponds to a translation of the bubble.
For quadrupole and higher deformations, however, $\delta F$ is always
negative, showing instability against arbitrary changes of shape.

\vglue 0.8cm
{\bf\noindent 3. Discussion}
\vglue 0.4cm

We have here considered only static properties of the quark--hadron interface.
Dynamical processes in the course of the phase transition will
destabilize the interface even more.
For example, the necessity of carrying heat and baryon number away
from a growing vacuum bubble will lead to a negative effective surface
tension and hence to complex phenomena of pattern formation.
On the other hand, interaction of bubbles, which was completely ignored in all
the above treatments, may play a crucial role in stabilizing the bubbles'
surface.

A word of caution: the above analysis relies on a bag model calculation$^3$ .
And although Monte Carlo measurements confirmed the smalleness of $\sigma$ as
well as the sign and magnitude of $\alpha$ of the bag model calculation, it is
still far from actually confirming even qualitatively the appearance of a
minimum in $F(R)$ for a hadron bubble with finite radius surrounded by the
quark gluon plasma, as depicted in fig. 2. As a matter of fact, the lattice
results show that the interface is about 2fm thick, and it is clearly not
negligible compared to the temperature or to $R_0$ , whereas  the stability
analysis is based on a thin wall approximation.

\vskip 24pt
{\bf \noindent 4. Acknowledgements \hfil}
\vglue 0.4cm
This work was supported by a Wolfson Research Award administered by the
Israel Academy of Sciences and Humanities.
\vglue 0.6cm
{\bf\noindent 5. References \hfil}
\vglue 0.4cm


\begin{thebibliography}{9}

\def\PRD#1{Phys. Rev. D {\bf #1}}
\def\PLB#1{Phys. Lett. B {\bf #1}}
\def\PRC#1{Phys. Rev. C {\bf #1}}
\def\ZPC#1{Z. Phys. C {\bf #1}}
\def\NPA#1{Nucl. Phys. {\bf A#1}}
\def\NPB#1{Nucl. Phys. {\bf B#1}}
\def\AP#1{Ann. Phys. (NY) {\bf #1}}
\def\ibid#1{{\it ibid.\/} {\bf #1}}


\bibitem{FJ}E.~Farhi and R.~L.~Jaffe, \PRD{30} (1984) 2379.
\bibitem{K1}K. Kajantie and L. K\"arkk\"ainen, \PLB{214} (1988) 595;
       K. Kajantie, L. K\"arkk\"ainen, and K.~Rummukainen, \NPB{333}
(1990) 100. \newline
       S.~Huang, J.~Potvin, C.~Rebbi, and S.~Sanielevici, \PRD{42} (1990)
	2864; {\bf 43} (1991) 2056(E).\newline
       R.~Brower, S.~Huang, J.~Potvin, and C.~Rebbi, Boston University
preprint BUHEP-92-3 (unpublished, 1992).
\bibitem{MS}I.~Mardor and B.~Svetitsky, \PRD{44} (1991) 878.
\bibitem{BS}B. Svetitsky,
lecture given at the Workshop on QCD at Finite Temperature and Density,
Brookhaven National Laboratory, August 1991, Tel Aviv preprint TAUP 1909--91
(1991), to appear in proceedings.
\bibitem{K2}K. Kajantie, L. K\"arkk\"ainen, and K.~Rummukainen, Helsinki
	preprint HU-TFT-92-1 (unpublished, 1992).
\bibitem{BB}R.~Balian and C.~Bloch, \AP{60} (1970) 401.
\bibitem{BB} J. Potvin, these Proceedings.

\end{thebibliography}
\end{document}